\documentclass[english,prstab, twocolumn]{revtex4}
\usepackage[T1]{fontenc}
\usepackage[latin9]{inputenc}
\usepackage{amsmath}
\usepackage{amssymb}
\usepackage{graphicx}

\makeatletter
\@ifundefined{textcolor}{}
{%
 \definecolor{BLACK}{gray}{0}
 \definecolor{WHITE}{gray}{1}
 \definecolor{RED}{rgb}{1,0,0}
 \definecolor{GREEN}{rgb}{0,1,0}
 \definecolor{BLUE}{rgb}{0,0,1}
 \definecolor{CYAN}{cmyk}{1,0,0,0}
 \definecolor{MAGENTA}{cmyk}{0,1,0,0}
 \definecolor{YELLOW}{cmyk}{0,0,1,0}
}

\makeatother

\usepackage{babel}
\begin{document}

\title{Beam-beam Effects of 'Gear-changing' in Ring-Ring Colliders}

\author{Yue Hao}

\author{V.N. Litvinenko}

\author{V. Ptitsyn}

\affiliation{Brookhaven National Laboratory, Upton, NY 11973}

\affiliation{Department of Physics, StonyBrook University, Stony Brook, NY 11794}
\begin{abstract}
In ring-ring colliders, the collision frequency determines the bunch
structures, e.g. the time between the bunches in both rings should
be identical. Because of relatively low relativistic speed of the
hadron beam in sub-TeV hadron-hadron- and electron-ions-colliders,
scanning the hadron beam's energy would require either a change in
the circumference of one of the rings, or a switching of the bunch
(harmonic) number in a ring. The later would cause so-called 'gear-changing',
i.e. the change of the colliding bunches turn by turn. In this article,
we study the difficulties in beam dynamics in this 'gear-changing'
scheme.
\end{abstract}


\maketitle

\section{Introduction}

In sub-TeV hadron colliders\cite{eg_rhic} and the ring-ring scheme
for electron-ion colliders (EICs)\cite{eg_elic}, scanning the energy
of one of the hadron beams is desired. This need brings the challenge
of synchronizing the two colliding beams. Since the hadron beams are
not ultra-relativistic in the sub-TeV energy range, their velocity
is significantly dependent upon their energy. To assure that both
beams collide at a fixed interaction point (IP), the circumference
of hadron ring must be adjusted when its energy changes. For instance,
if one changes the hadron beam from 250 GeV per nucleon (GeV/n) to
25 GeV/n, its velocity changes about 0.07\%. Synchronizing the collision
requires a proportional change of the ring's circumference. For a
3km circumference, such an energy change corresponds to a change in
pass length of about 2.1 meters. Altering the circumference would
necessitate mechanically moving a large part (an arc) of the ring,
so representing a major cost and operational obstacle. 

To reduce or avoid such an adjustment, one can choose to change the
RF harmonic number, \textit{h}, of the ring, i.e. to change the number
of bunches in the ring. This could allow the hadron beam to scan a
discrete set of energies without adjusting the circumference. For
the same example of a nominal energy 250 GeV/n, where the RF harmonic
number is 2500, one can adjust the hadron beam's energy to 35.1 GeV/n,
24.9 GeV/n and 20.4 GeV/n by changing the harmonic number of either
of the ring by 1, 2 and 3 units, respectively, without changing the
ring's circumference. If energies between those discrete values are
required, a smaller adjustment of the circumference still would be
required. In this example, one has to adjust the circumference of
the ring by 1.05 meters, for the entire energy range between 250 GeV/n
and 35.1 GeV/n. If the rings' circumferences are equal and the RF
harmonic numbers are not, one RF bucket in one of them will overlap
with multiple RF buckets in the other ring in successive collision.
Therefore the bunch is one ring will collide with multiple bunches
in the other ring in different turns of revolution. 

This asymmetric collision pattern could introduce complications with
the beam dynamics. In this paper, we concentrate exploring the consequence
of these asymmetric patterns in two rings, including a multi-bunch
offset/dipole instability (section \ref{sec:dipole moment}) and a
multi-bunch beam-size/quadrupole instability(section \ref{sec:quad_moment})
due to the 'gear-changing' beam-beam collision pattern, as well as
the single particle dynamics at presence of abort gap in the opposing
bunch train(section \ref{sec:bunch_gap}).

\section{Multi-bunch, dipole moment analysis\label{sec:dipole moment}}

One possible side effect of this asymmetric collision pattern is the
instability of the centroid of both beams. This topic is first studied
in \cite{barycentre_motion}. In this section, we present an alternative
matrix method to study this problem; it easily can be extended to
the case of large bunch numbers in both rings.

We will consider two rings of collider that have different bunch numbers,
$N_{1}$ and $N_{2}$ ($N_{2}>N_{1}$), respectively. They are evenly
distributed in the corresponding RF buckets. No bunch gap is considered
herein. The time separations between the bunches are same in both
rings, so that the bunches are synchronized to collide at the IP.
In this case, one bunch will collide not only with a single bunch
in the opposing ring, but with $N_{c}$ bunches successively, where
$N_{c}=LCM\left(N_{1},N_{2}\right)/N_{1}$. $LCM\left(\right)$ denotes
the least common multiple of the arguments. If $N_{1}$ and $N_{2}$
are relatively prime numbers, the bunch will collide with all bunches
of the opposing beam. We also assume that the optics function at IP
of the two rings are same and the two colliding beams have same emittance
and intensity. Therefore the beam-beam parameters, $\xi$, for all
bunches are same.

\begin{figure*}
\includegraphics[width=1\columnwidth]{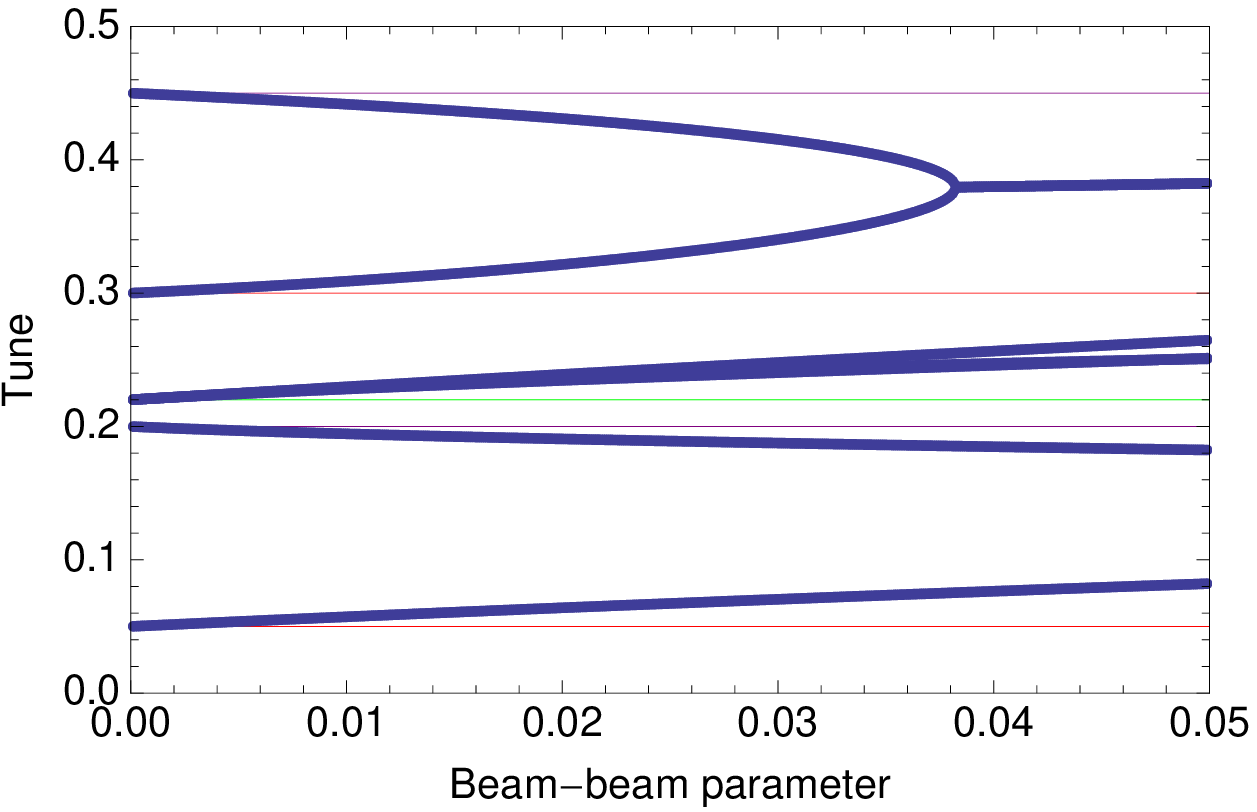}\includegraphics[width=1\columnwidth]{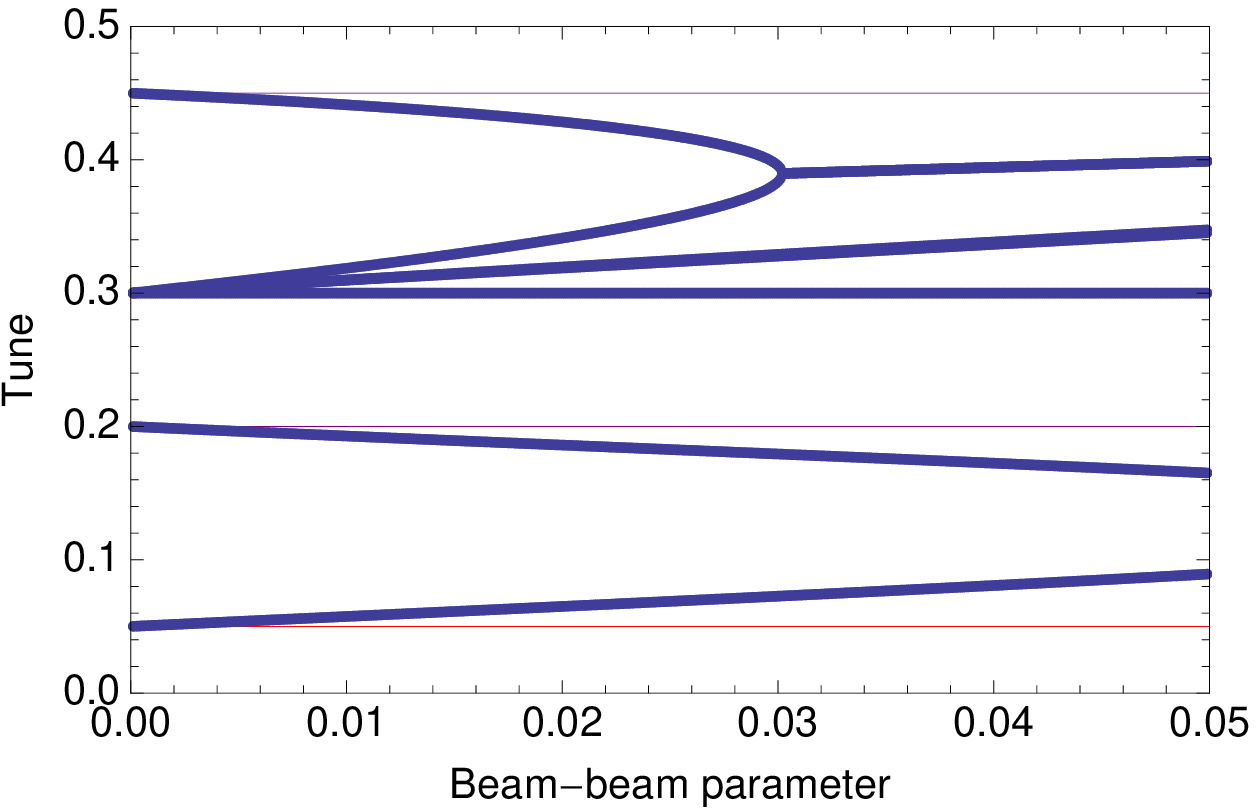}

\includegraphics[width=1\columnwidth]{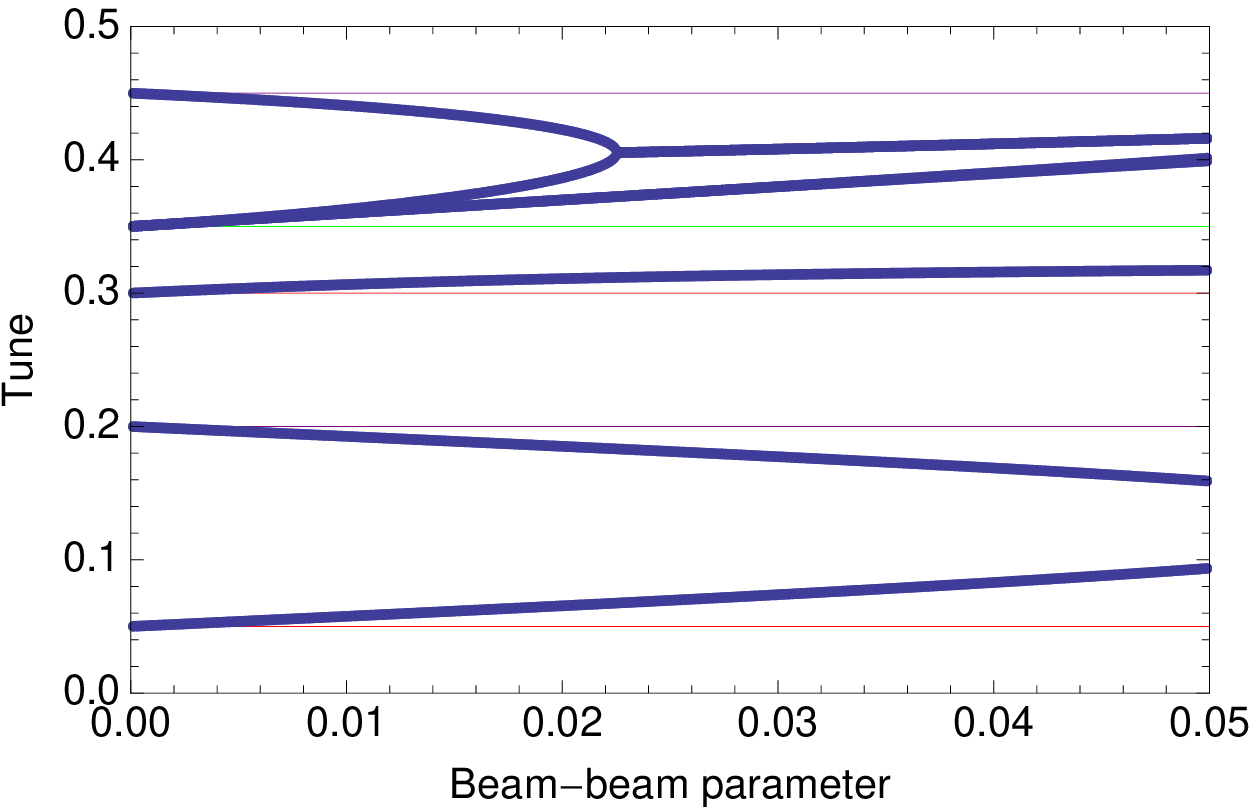}\includegraphics[width=1\columnwidth]{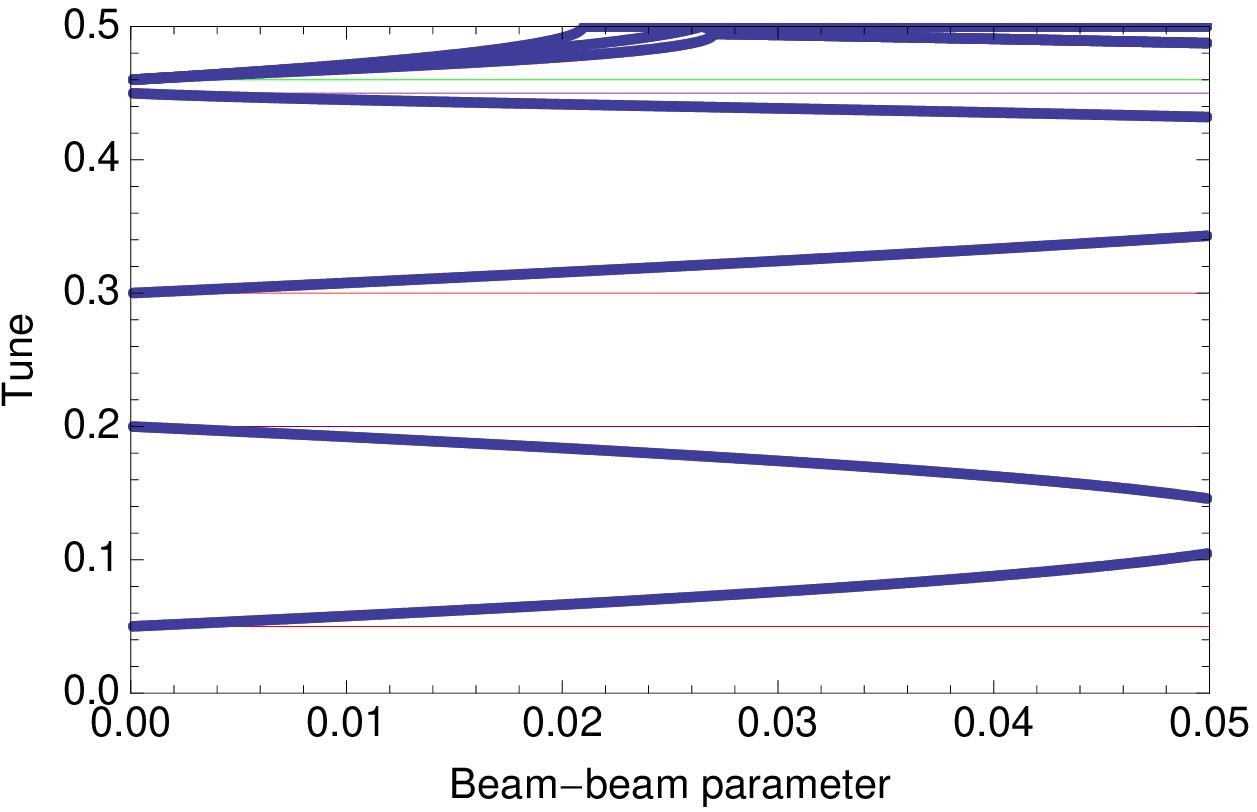}

\caption{The tunes extracted from matrix $M_{T}$ with different beam-beam
parameters $\xi$. The vertical grid lines represent the tunes $\Xi_{i}$
when there is no beam-beam interaction. The green one denotes $\Xi_{1}$
and the red and purple ones represent the $\Xi_{2}$ or $1-\Xi_{2}$
in the range {[}0, 0.5{]} respectively. All figures have the number
of bunches in two ring as $\left(N_{1},N_{2}\right)=\left(3,4\right)$
with different tunes. The tunes are the following: Top left $\left(\nu_{1},\nu_{2}\right)=\left(0.22,0.4\right)$;
top right $\left(\nu_{1},\nu_{2}\right)=\left(0.3,0.4\right)$; bottom
left $\left(\nu_{1},\nu_{2}\right)=\left(0.35,0.4\right)$ and bottom
right $\left(\nu_{1},\nu_{2}\right)=\left(0.46,0.4\right)$. \label{fig:dipole_tune_merge}}
\end{figure*}

The dipole moments of each bunch in ring 1 will affect the $N_{c}$
bunches in the other rings within the $N_{c}$ turns. The dipole moments
must be stable in these collision processes. To characterize the dipole
moment dynamics of $N_{1}+N_{2}$ bunches, we need to establish an
equivalent 'one-turn' matrix of a repetitive pattern for our analysis.
The matrix is a $2(N_{1}+N_{2})$ square matrix to denote the centroid
motion of all the bunches. Accordingly, we have to calculate the matrix
for successive $LCM\left(N_{1},N_{2}\right)$ collisions that corresponds
to $LCM(N1,N2)/N_{1}$ turns for the beam 1; meanwhile beam 2 finishes
$LCM(N1,N2)/N_{2}$ turns. When $N_{1}$ and $N_{2}$ are large and
relative primes, it is very time consuming to construct this matrix
for the repetitive pattern.

However, we found that we could simplify the calculation via the cyclic
permutation matrix $P_{T}$ (Eq. (\ref{eq:def_p})) that enables us
readily to establish such a 'one-turn' matrix. The effect $2(N_{1}+N_{2})\times2(N_{1}+N_{2})$
'one turn' total matrix $M_{T}$ has the form:
\begin{equation}
M_{T}=P_{T}M_{k}M_{\beta}
\end{equation}
in which, the betatron oscillation matrix $M_{\beta}$ is constructed
as follows:
\begin{equation}
M=\begin{pmatrix}I\left(N_{1}\right)\otimes M\left(\theta_{1}\right) & 0 & 0\\
0 & I\left(N_{1}\right)\otimes M\left(\theta_{2}\right) & 0\\
0 & 0 & I\left(2N_{2}-2N_{1}\right)
\end{pmatrix}
\end{equation}
Here, $M\left(\theta\right)$ is the $2\times2$ one turn matrix at
IP of both rings with $\theta_{1}$ and $\theta_{2}$, respectively,
as the one turn phase advance of two rings respectively, symbol $ $$\otimes$
denotes the matrix outer product. 

$M_{k}$ is the linearized beam-beam kick of the $i^{th}$ bunch ($i=1\ldots N_{1}$)
of beam 1 with $i{}^{th}$ one of beam 2. $M_{k}$ can be established
by changing the following elements from a unit matrix,

\begin{equation}
\begin{array}{ccc}
M_{k}\left[2i,2i-1\right] & = & 1/f_{bb}\\
M_{k}\left[2(N_{1}+i),2(N_{1}+i)-1\right] & = & 1/f_{bb}\\
M_{k}\left[2i,2(N_{1}+i)-1\right] & = & -1/f_{bb}\\
M_{k}\left[2(N_{1}+i),2i-1\right] & = & -1/f_{bb}
\end{array}
\end{equation}
with the focal length of the beam-beam interaction defined as $1/f_{bb}=\pm4\pi\xi/\beta$,
where the positive sign corresponds to the same sign of charge of
the two colliding particles. The structure of matrix $M_{\beta}$
and $M_{k}$ shows that the betatron oscillation and the beam-beam
kick apply to all the bunches in beam 1, but only the first $N_{1}$
bunches of beam 2, which collide with all bunches with their counterparts
in beam 1. 

The rotation matrix $P_{T}$ is constructed as follows:
\begin{equation}
P_{T}=\begin{pmatrix}I\left(2N_{1}\right) & 0\\
0 & P^{N_{2}-N_{1}}\left(N_{2}\right)\otimes I\left(2\right)
\end{pmatrix}\label{eq:def_pt}
\end{equation}
where $I\left(n\right)$ and $P\left(n\right)$, respectively, are
the identity matrix and the cyclic permutation matrix of rank $n$.
The permutation matrix $P$ reads

\begin{equation}
P=\begin{pmatrix}0 & 0 & \ldots & 0 & 1\\
1 & 0 & \ldots & 0 & 0\\
0 & \ddots & \ddots & \vdots & \vdots\\
\vdots & \ddots & \ddots & 0 & 0\\
0 & \ldots & 0 & 1 & 0
\end{pmatrix}\label{eq:def_p}
\end{equation}
where the eigenvalues of $P$ with rank $k$ are $e^{i2m\pi/k}$,
$m=0,1,\cdots,k-1$. The rotation matrix $P_{T}$ will rotate the
beam 2 so that the remaining $N_{2}-N_{1}$ bunches will interact,
in turn, with the opposing bunches in beam 1. This method avoids our
having to construct a complicated matrix for the $LCM\left(N_{1},N_{2}\right)$
collisions without any approximation.

We first discuss the case without collision, i.e., the matrix $M_{k}$
is an identity matrix. The eigenvalues of the matrix $M_{T}$ has
$N_{1}$ degenerate pairs of $e^{\pm i\theta_{1}}$ that come from
the tune of ring 1 ($\theta_{1}=2\pi\nu_{1}$). However the eigenvalues
contributed from ring 2 has unusual form that reads $e^{\pm i\theta_{2j}}$,
$\theta_{2j}=2\pi(N_{1}\nu_{2}+j)/N_{2}$, $j=0,GCD(N_{1},N_{2}),\cdots,N_{2}$,
where $GCD(N_{1},N_{2})$ is the greatest common divider of two bunch
numbers. As expected, the absolute values of all eigenvalues are one.
Hereafter, we limit our discussion in the cases when $N_{1}$ and
$N_{2}$ are relative prime numbers.

$ $The effective tunes extracted from the total matrix $M_{T}$ are
defined as $\Xi_{i}$ ($i=1,\cdots,N_{1}+N_{2}$ ). When the beam-beam
parameter is zero, there are $N_{1}$ degenerate tunes $\Xi_{1,i}=\nu_{1}$
($i=1,\cdots,N_{1}$ ) from the tune of ring 1, and $N_{2}$ distinct
tunes $\Xi_{2,j}=(N_{1}\nu_{2}+j)/N_{2}$ ($i=1,\cdots,N_{2}$ ) from
ring 2. Including the beam-beam interaction is included, $\Xi_{1}$
and the $\Xi_{2j}$ will shift according to the beam-beam parameter.
The system becomes unstable when integer or half integer resonances
happens and one or more eigenvalues of the matrix $M_{T}$ have absolute
value larger than one. However, the resonance condition in this asymmetric
collision case is more complicated. We also note that, in this asymmetric
scheme, there exist $\pi$ modes similar to the symmetric collision
case. The $\pi$ modes occurs when $\Xi_{1}(\xi=0)=\Xi_{2,j}\left(\xi=0\right)$
. The tune of this mode remains intact with beam-beam interaction
of any strength. 

Figure \ref{fig:dipole_tune_merge} illustrates the instability of
the total map $M_{T}$, occurring when two eigen-tunes merge. In this
example, the attractive beam-beam force is used which corresponds
to the case of the electron-ion colliders. We choose the number of
bunches 3 and 4, respectively, in the two rings and fixed the tune
of the second ring as $\nu_{2}=0.4$. Therefore the effective tune
$\Xi_{2}=\left\{ 0.3,0.55,0.8,0.05\right\} $ when the beam-beam parameter
is zero, and the tune of ring 1 $\Xi_{1}=\nu_{1}$ is varied to explore
the structure of the tune map. In the top left figure, where $\nu_{1}=0.22$,
the instability happens when the two tunes of $\Xi_{2,1}=0.3$ and
$\Xi_{2,2}=0.55$ merges at beam-beam parameter $\xi\sim0.036$, which
is referred as the sum resonance $\Xi_{2,1}+\Xi_{2,2}\sim1$, i.e.
the half integer resonance of $N_{1}\nu_{2}$. The top right figure
illustrates the appearance of the $\pi$ mode when $\nu_{1}=0.3$
was chosen. We can observe that this tune remain intact when the beam-beam
parameter increases. The bottom left figure reflects the sum resonance
of $\Xi_{1}+\Xi_{2,2}\sim1$, which corresponds to the integer number
resonance of $N_{1}\nu_{2}+N_{2}\nu_{1}$, when $\nu_{1}$ is set
to 0.35. A trivial case of half integer resonance at $\Xi_{1}\sim0.5$
is shown in the bottom right one with $\nu_{1}=0.46$. 

These resonances are based on the integer turns of ring 1. Therefore
only the half integer resonance of $N_{1}\nu_{2}$ and $\nu_{1}$
are observed directly. If we rewrite the total matrix based on the
integer turns of ring 2, the other two symmetric counterparts $N_{2}\nu_{1}$
and $\nu_{2}$ can be found. 

\begin{figure}
\includegraphics[width=1\columnwidth]{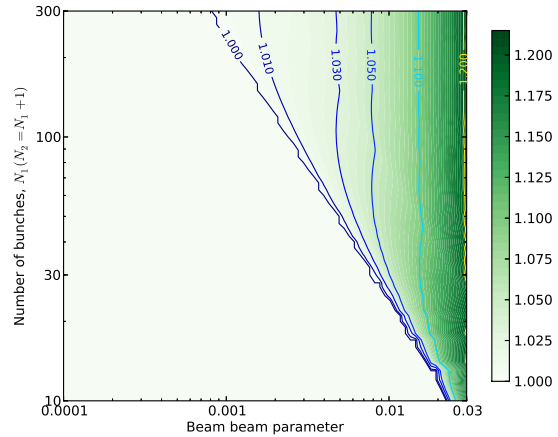}

\caption{The largest module of the eigenvalues of $M_{T}$ for different beam-beam
parameters and number of bunch in ring one $N_{1}$, while $N_{2}=N_{1}+1$.
Both tunes in two rings are optimized around 0.2 for each $N_{1}$.\label{fig:threshold_color_map}}
\end{figure}

From the above discussion, we learned that, in this asymmetric collision
pattern, there are many more linear resonances to be avoided than
in a symmetric pattern for any selected working point in the tune
diagram. The sum resonances may occurs between the $\Xi_{1}$ and
$1-\Xi_{2,j}$ or $\Xi_{2,i}$ and $1-\Xi_{2,j}$. The spacing between
$\Xi_{2}$ is $1/N_{2}$. If $N_{2}$ is a large number, it will strongly
limit the choice of the tunes in both rings and the maximum beam-beam
parameter, so to avoid linear resonances. In the linear beam-beam
tune shift approximation, $d\Xi/d\xi=1$, therefore we have to limit
the beam-beam tune shift to $\xi\lesssim1/2N_{2}$, as well as to
select proper tunes to avoid the sum resonance. The working point
in storage rings is determined by many other factors, such as the
nonlinear resonances, dynamic aperture optimization, as well as spin
resonances for polarized beam colliders. Our study indicates that
a fine-tuning of the working points is also required to avoid a dense
net of resonance lines separated by approximately $1/N_{2}$. In figure
\ref{fig:threshold_color_map}, we vary $N_{1}$ and beam-beam parameter
$\xi$ in a large ranges and optimize both tunes of the rings, $\nu_{1}$
and $\nu_{2}$, around 0.2 to minimized the largest modulus of the
eigenvalue of the matrix $M_{T}$ for an individual $N_{1}$. In these
processes, we fixed $N_{2}=N_{1}+1$. The color map and the contour
lines in figure \ref{fig:threshold_color_map} indicates optimized
largest modulus values, and therefore the stable region is defined
by the contour line of modulus equaling 1. As expected the boundary
of the stable region is a linear line in the logarithmic scale, which
indicates $\xi_{max}\sim1/N_{1}\sim1/N_{2}$. The slope of this line
is about 0.25, as derived by Linear regression, therefore, the stability
boundary reads $\xi_{max}\approx1/4N_{1}$. For bunch numbers of $N_{1}=100$
and $N_{2}=101$, the maximum beam-beam parameter is only $2.5\times10^{-3}$,
which is much smaller than the beam-beam parameter realized in both
hadron and lepton colliders in a symmetric collision pattern. 

To restore the beam-beam parameters of the asymmetric collision pattern
to the symmetric one, a bunch-by-bunch transverse feedback system
(transverse damper) is required. The largest modulus plotted in figure
\ref{fig:threshold_color_map} helps to determine the damping time
of such damper. To restore the limit of the typical beam-beam parameter
in a hadron collider ($\xi\sim0.02$), the damping time of such damper
must be less than 10 turns for a large $N_{1}$ ($\geq100$), since
the largest modulus is around 1.1 for those bunch numbers. The radiation
damping of the electron beam are not helpful, since the damping time
is much longer than the requirement. The damping time of \textasciitilde{}10
turns is very challenging requirement and, if it is possible, would
require an ultra low-noise pick-up and kicker system in either the
hadron or the electron rings.

\section{Multi bunch, quadrupole moment analysis\label{sec:quad_moment}}

In this section, we detail our findings from extending our study to
the transverse beam size instability in this 'gear-changing' collision
pattern. Herein, we assume that the dipole offset of all bunches are
damped to zero. 

We are interested in the vector of second order momentum $V=\left(\langle x^{2}\rangle,-\langle xx\prime\rangle,\langle x\prime^{2}\rangle\right)^{T}$
and its propagation under a linearized beam-beam force. In this case,
the emittance of the beam is constant. Therefore, for a $2\times2$
betatron transfer map $M$, the transport matrix $R$ for the vector
$V$ is well know. 

\begin{equation}
R\left(M\right)=\begin{pmatrix}M_{11}^{2} & -2M_{11}M_{12} & M_{12}^{2}\\
-M_{11}M_{21} & M_{11}M_{22}+M_{12}M_{21} & -M_{12}M_{22}\\
M_{21}^{2} & -2M_{21}M_{22} & M_{22}^{2}
\end{pmatrix}
\end{equation}
The one-turn transport matrix for second momentum including linearized
beam-beam effect is 
\begin{equation}
R\left(K\right)\cdot R\left(M_{0}\right)
\end{equation}
 where
\[
M_{o}=\begin{pmatrix}\cos\mu_{\beta} & \beta^{*}\sin\mu_{\beta}\\
-\sin\mu_{\beta}/\beta^{*} & \cos\mu_{\beta}
\end{pmatrix}\mbox{ and }K=\begin{pmatrix}1 & 0\\
1/f_{bb} & 1
\end{pmatrix}
\]
where $\mu_{\beta}$ is the transverse betatron tune and $f_{bb}$
is the focal length of the linearized beam-beam force. The matrix
$R\left(M_{0}\right)$ has three eigenvalues as $\exp\left[\pm2i\mu_{\beta}\right]$
and 1. The matrix $R\left(K\right)$ has the explicit form:
\begin{equation}
R\left(K\right)=\begin{pmatrix}1 & 0 & 0\\
-1/f & 1 & 0\\
1/f^{2} & -2/f & 1
\end{pmatrix}
\end{equation}
We note that $f_{bb}$ is inversely proportional to the quadrate of
rms beam size of the opposing beam provided that the beam is round.
Therefore we encounter the difficulty that the two beams' size does
not has a linear cross-talk as in the case of centroid offset case,
described the previous section.

We may circumvent this problem by studying the evolution of infinitesimal
second order momentum offset from the unperturbed vector. We may write
$V=V_{0}+V$, where $V_{0}$ is unperturbed vector $V_{0}=\left(\langle x^{2}\rangle_{0},-\langle xx\prime\rangle_{0},\langle x\prime^{2}\rangle_{0}\right)^{T}$
and $V_{1}=\left(d\langle x^{2}\rangle,-d\langle xx\prime\rangle,d\langle x\prime^{2}\rangle\right)^{T}$
is the its first order deviation. The unperturbed vector $V_{0}$
is related to the dynamic optics function as $V_{0}=\left(\beta,\alpha,\gamma\right)_{d}\varepsilon$,
where $\varepsilon$ is the beam emittance. They are the ones that
deviate from the design optics at IP due to the beam-beam force. If
we assume the betatron phase advance is different in the two colliding
rings, the dynamics optics functions also differ. Using $x$ and $y$
to denote two colliding bunches, we can write the transformation of
the linearized beam-beam interaction as follows:
\begin{eqnarray*}
d\langle x^{2}\rangle_{bb} & = & d\langle x^{2}\rangle\\
-d\langle xx\prime\rangle_{bb} & = & -\frac{d\langle x^{2}\rangle}{f}+\left(-d\langle xx\prime\rangle\right)+\frac{\beta_{x}}{f\beta_{y}}d\langle y^{2}\rangle\\
d\langle x\prime^{2}\rangle_{bb} & = & \frac{d\langle x^{2}\rangle}{f^{2}}-2\frac{-d\langle xx\prime\rangle}{f}+d\langle x\prime^{2}\rangle\\
 &  & -2\frac{\alpha_{x}+\beta_{x}/f}{f\beta_{y}}d\langle y^{2}\rangle
\end{eqnarray*}
where the subscript $bb$ represents the deviation vector after the
beam-beam interaction and the optics functions are the dynamics ones.

Therefore, we similarly can build the 'one-turn map' $M_{T}$ of the
perturbed second order momentum as detailed in the previous section,
$M_{T}=P_{T}M_{k}M_{\beta}$. We reuse all the symbols, however, it
is straightforward that all matrices are $3\left(N_{1}+N_{2}\right)$
rank square matrices. $M_{\beta}$ is built from $R\left(M\right)$
by
\[
M_{\beta}=\begin{pmatrix}I_{N_{1}}\otimes R_{M}\left[\theta_{1}\right] & 0 & 0\\
0 & I_{N_{1}}\otimes R_{M}\left[\theta_{2}\right] & 0\\
0 & 0 & I_{3N_{2}-3N_{1}}
\end{pmatrix}
\]
where $I_{N}$ is the identity matrix with rank $N$, and $R_{M}\left[\theta\right]$
denotes $R\left[M_{0}\left(\theta\right)\right]$ for space saving.

The $M_{k}$ reads 
\[
M_{k}=\begin{pmatrix}I\left(N_{1}\right)\otimes R\left[K\right] & 0 & 0\\
0 & I\left(N_{1}\right)\otimes R\left[K\right] & 0\\
0 & 0 & I\left(3N_{2}-3N_{1}\right)
\end{pmatrix}
\]
with additional terms modified as
\begin{eqnarray*}
M_{k}[3i-1,3i+3N_{1}-2] & = & \frac{\beta_{x}}{f\beta_{y}}\\
M_{k}[3i,3i+3N_{1}-2] & = & -2\frac{\alpha_{x}+\beta_{x}/f}{f\beta_{y}}\\
M_{k}[3i+3N_{1}-1,3i+3N_{1}-2] & = & \frac{\beta_{y}}{f\beta_{x}}\\
M_{k}[3i+3N_{1},3i+3N_{1}-2] & = & -2\frac{\alpha_{y}+\beta_{y}/f}{f\beta_{x}}
\end{eqnarray*}
with the index $i=1,2,\cdots,N_{1}$. W can construct the $P_{T}$
as follows:
\begin{equation}
P_{T}=\begin{pmatrix}I\left(3N_{1}\right) & 0\\
0 & P^{N_{2}-N_{1}}\left(N_{2}\right)\otimes I\left(3\right)
\end{pmatrix}\label{eq:def_pt-quad}
\end{equation}
where the permutation matrix $P$ has the same definition (Eq. \ref{eq:def_p})
as given in the last section.

\begin{figure}
\includegraphics[width=1\columnwidth]{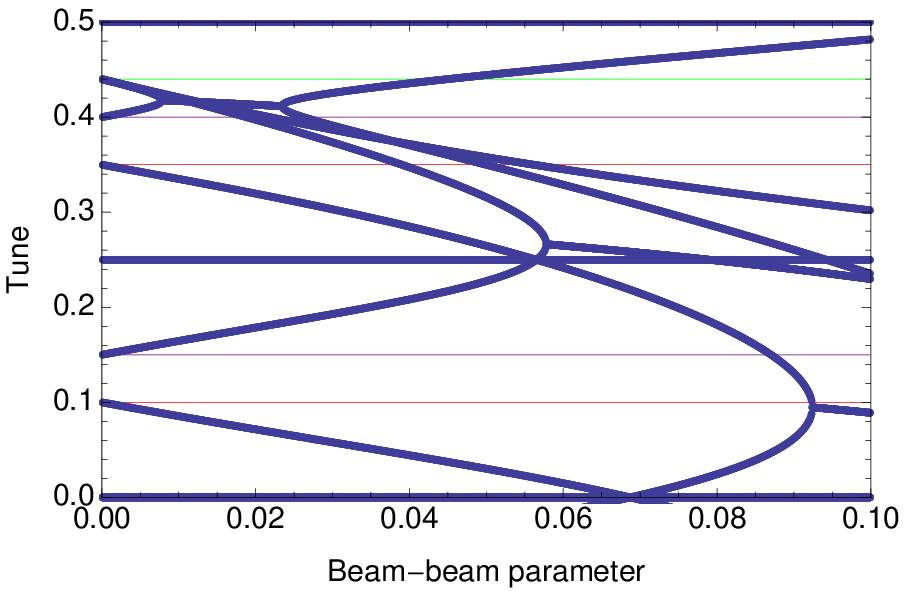}

\includegraphics[width=1\columnwidth]{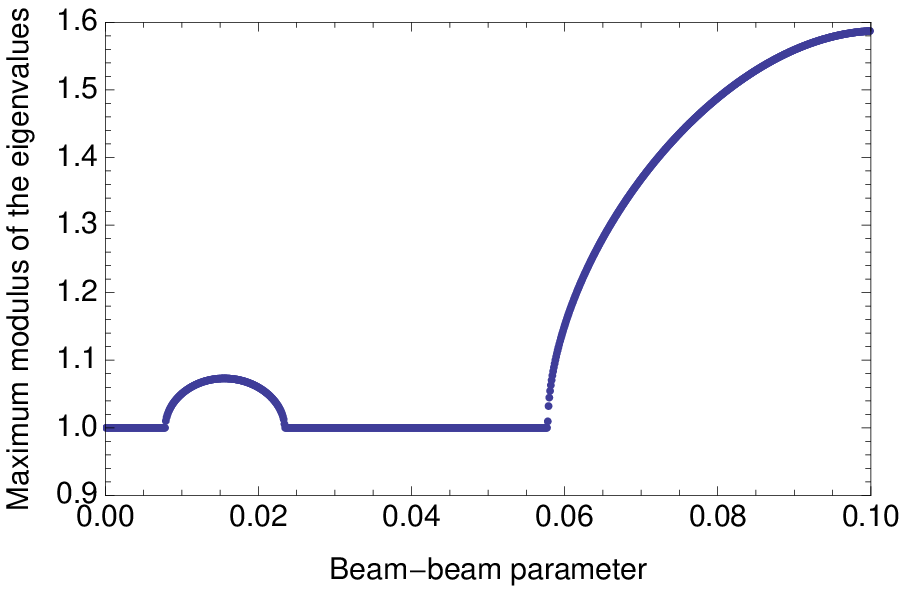}

\caption{The tunes extracted from second order momentum one-turn matrix $M_{T}$
(top) and the largest modulus of its eigenvalues (bottom) as function
of the beam-beam parameter. In this example, the number of bunches
in two rings are: $\left(N_{1},N_{2}\right)=\left(3,4\right)$. The
tunes are: $\left(\nu_{1},\nu_{2}\right)=\left(0.22,0.4\right)$.
\label{fig:The-tunes-extracted-quad}}
\end{figure}

The tunes of the second order momentum extracted from the matrix $M_{T}$
also have similar structures. Without beam-beam interactions, the
tunes has the contribution from the ring 1 as $\Xi_{1,i}=2\mu_{1}$
($i=1,2,\cdots,N_{1}$) and the ones from the ring 2 as $\Xi_{2,j}=2\mu_{2}N_{1}/N_{2}+j/N_{2}$,
($j=1,2,\cdots,N_{2}$). There also are 'dummy' tunes, which are $\Xi_{0}=j/N_{2}$,
($j=1,2,\cdots,N_{2}$). They are contributed from the eigenvalue
1 of the matrix $R(M_{0})$ and remain constant at non-zero beam-beam
interaction strength. When the beam-beam parameter is non-zero, the
resonance can occur at certain beam-beam parameters, which can be
predicted by the modulus and the arguments (tunes) of the eigenvalues
of $M_{T}$. Figure \ref{fig:The-tunes-extracted-quad} illustrates
the resonance conditions of the sum resonances of $2N_{2}\nu_{1}+2N_{1}\nu_{2}$.
We note that the stable region of the one-turn matrix $M_{T}$ may
not be continuous as the beam-beam parameter increases, as shown in
bottom graph of figure \ref{fig:The-tunes-extracted-quad}. 

The separation of $\Xi_{2}$ lines is $1/N_{2}$, which is the same
as the dipole analysis in last section. The difference is that the
beam-beam tune shift $d\Xi/d\xi$ is 2 instead of 1, in second order
momentum case, when beam-beam tune-shift is small. Therefore, maximum
achievable beam-beam parameter is expected to be smaller than in the
dipole case, before the first linear resonance is observed. However,
there may be another stable region followed by a unstable region as
show in \ref{fig:The-tunes-extracted-quad}. Therefore it is difficult
to conclude a simple criteria of stable system. 

\begin{figure}
\includegraphics[width=1\columnwidth]{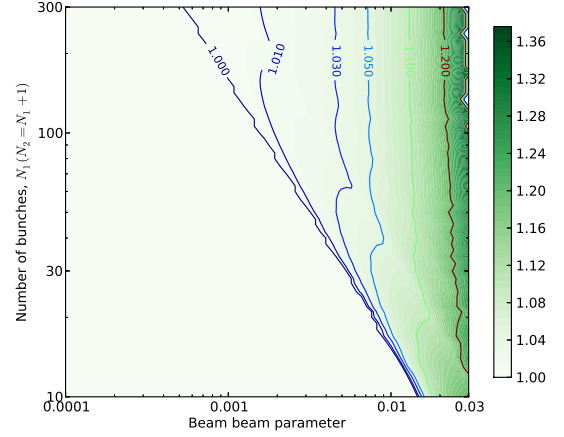}

\caption{The largest module of the eigenvalues of $M_{T}$ for different beam-beam
parameters and number of bunches in ring one $N_{1}$, while $N_{2}=N_{1}+1$,
for the second order momentum case. Both tunes in two rings are optimized
around 0.2 for each $N_{1}$.\label{fig:threshold_color_map-quad}}
\end{figure}

\begin{figure*}
\includegraphics[width=1\columnwidth]{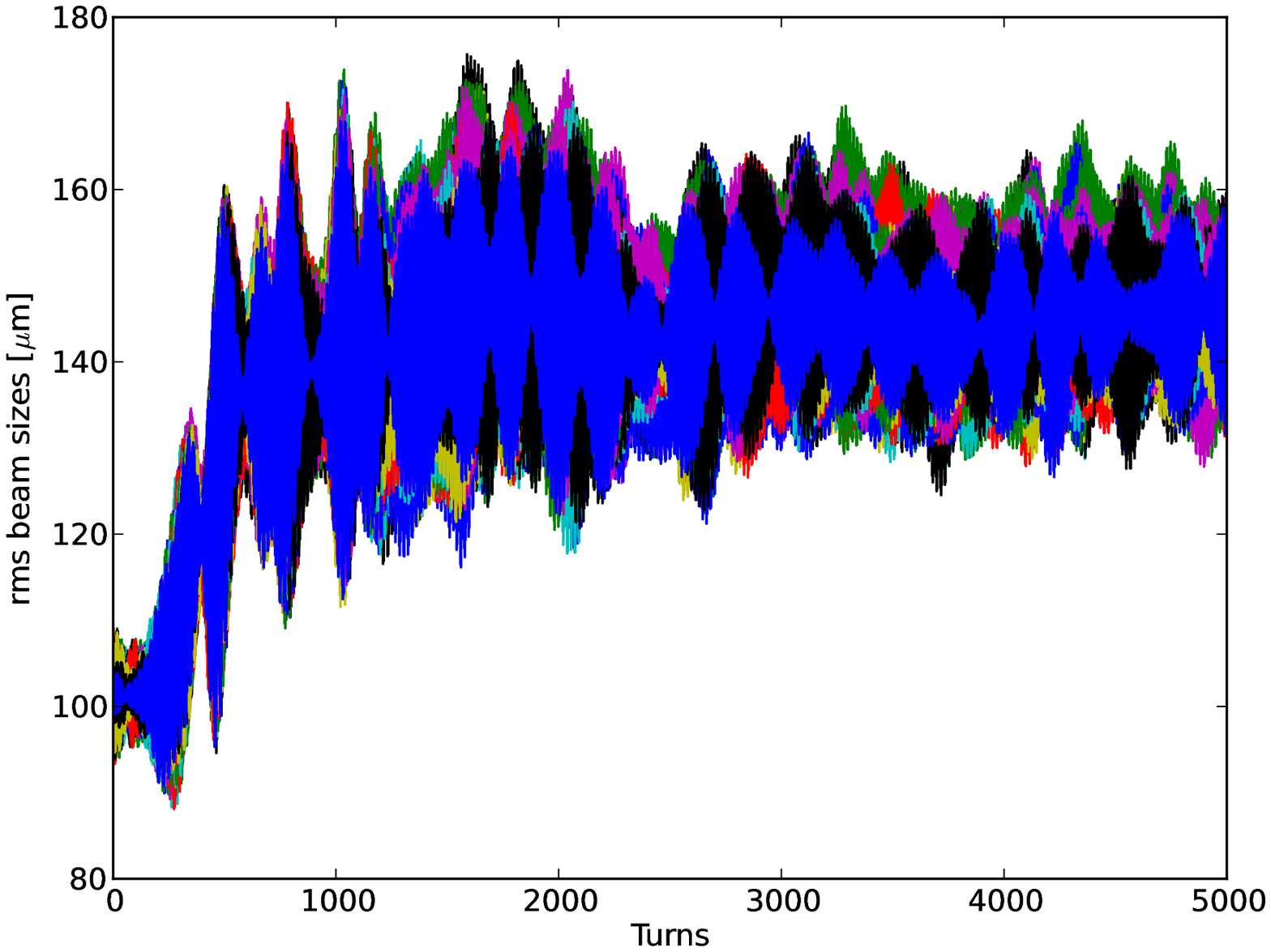}\includegraphics[width=1\columnwidth]{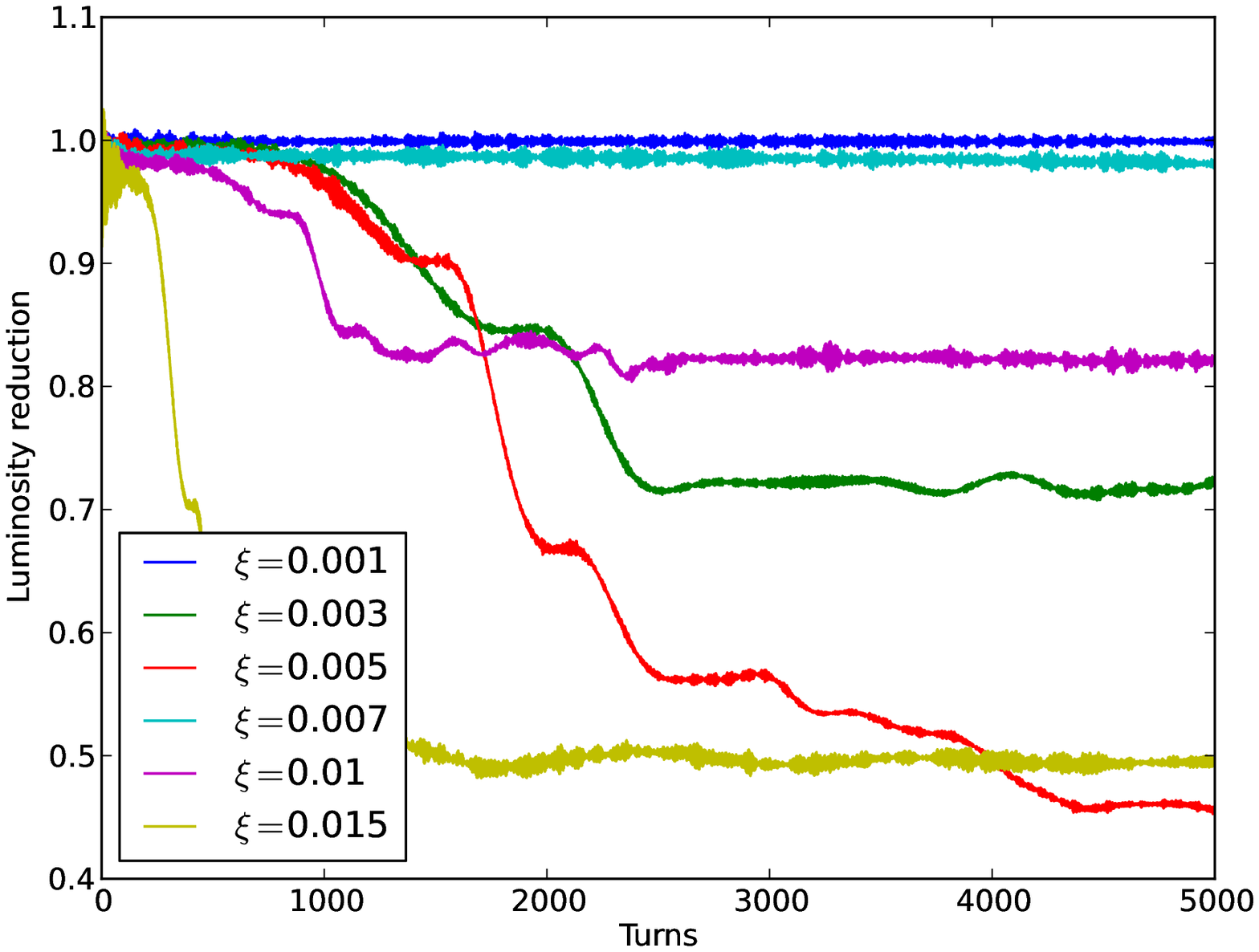}

\includegraphics[width=1\columnwidth]{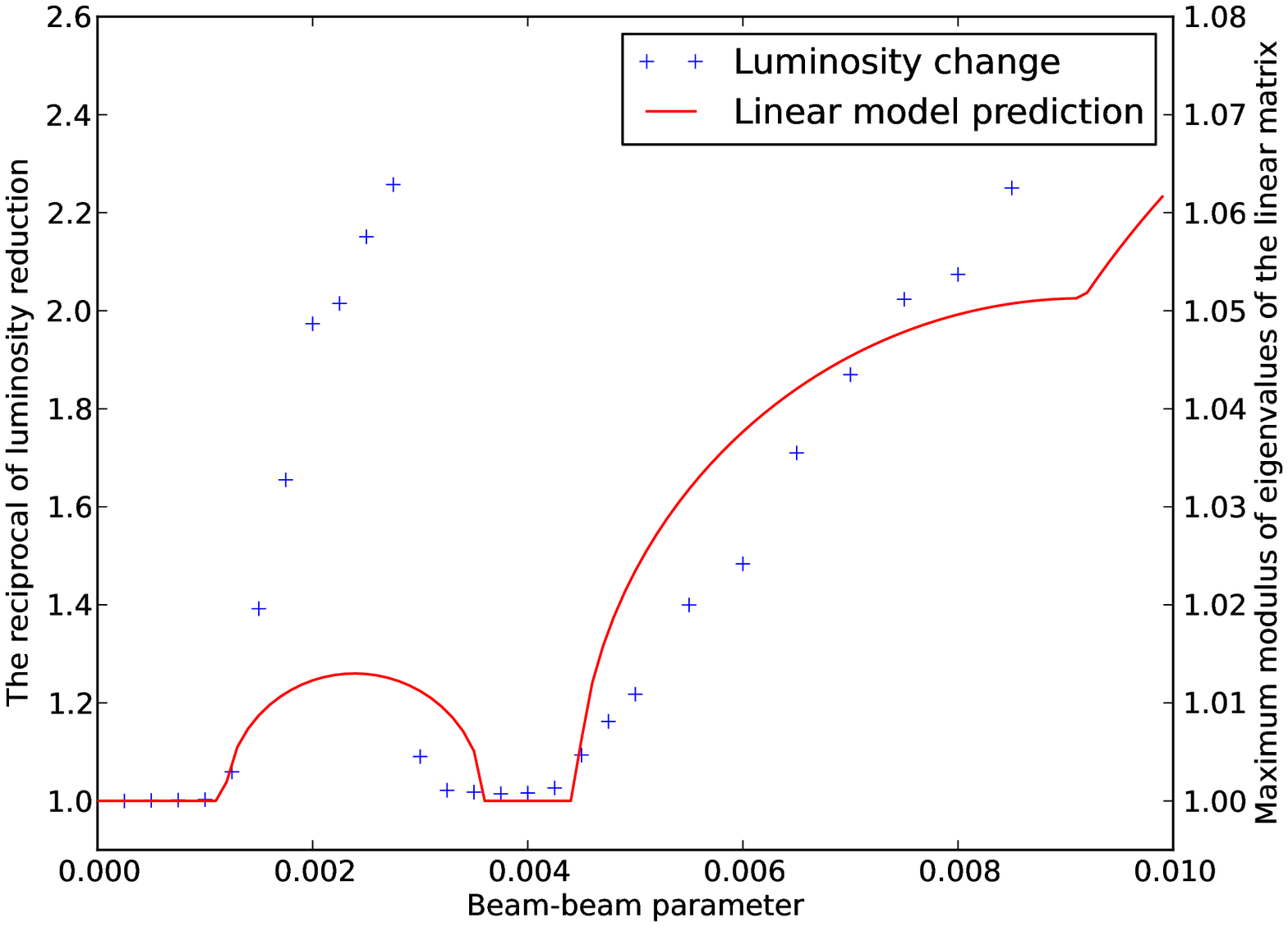}\includegraphics[width=1\columnwidth]{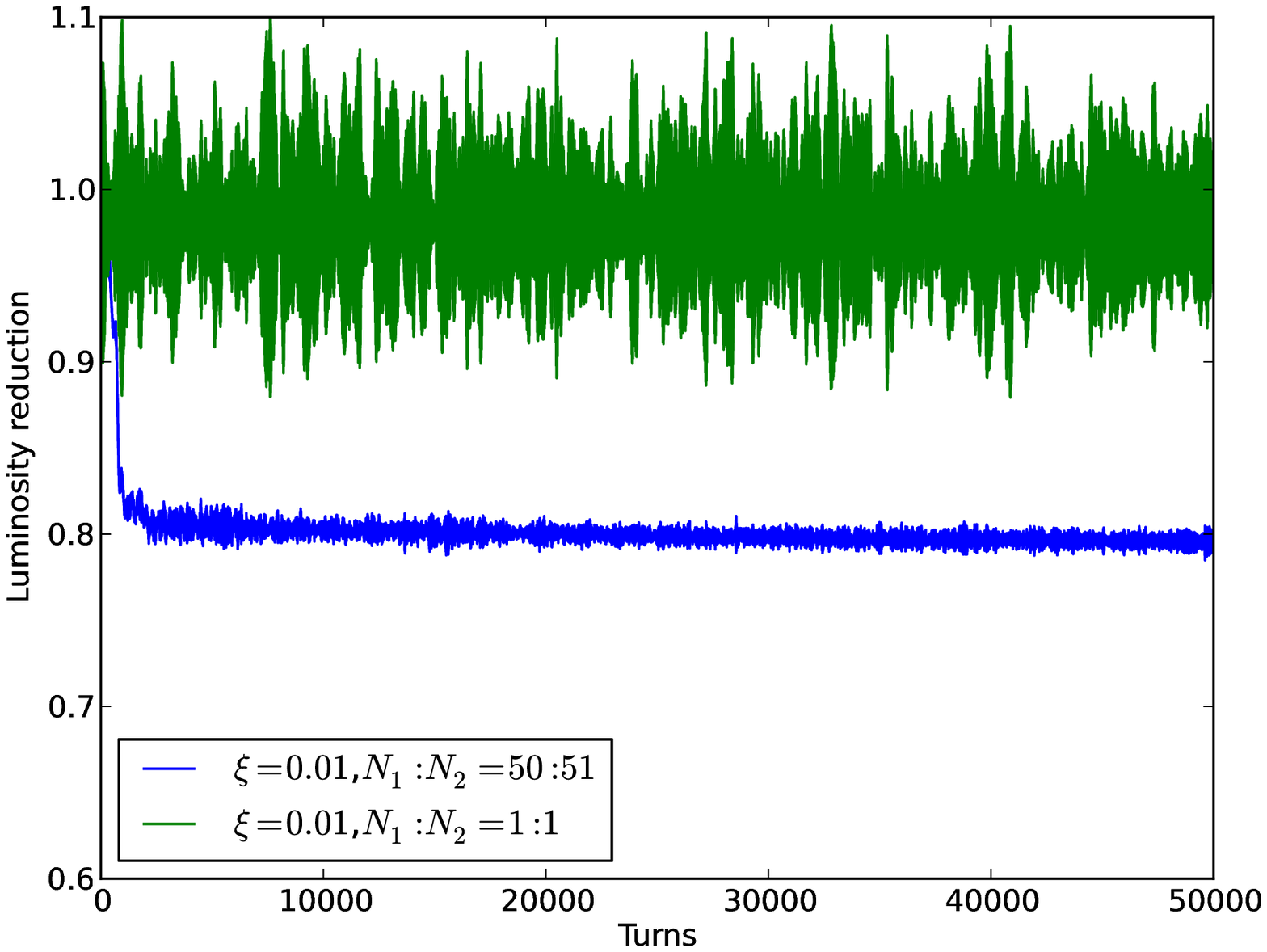}

\caption{The results of nonlinear 1-D beam-beam simulation for $\left(N_{1},N_{2}\right)=\left(50,51\right)$.
Top left: The beam size evolutions for $\xi=0.015$, which indicates
instability of the beam size. Top right: The luminosity reduction
due to the second order momentum instability in asymmetric collision
scheme. Bottom left: The reciprocal of luminosity reduction at the
end of 5000 turns and the comparison with linear matrix model. The
beam-beam parameter in the nonlinear simulation is scaled by 0.5 in
this plot. Bottom right: Comparison of the asymmetric and symmetric
collision scheme. \label{fig:quad_simulation} The tunes of two rings
are all set to 0.68 to avoid lower order nonlinear resonances.}
\end{figure*}

\begin{figure}
\includegraphics[width=1\columnwidth]{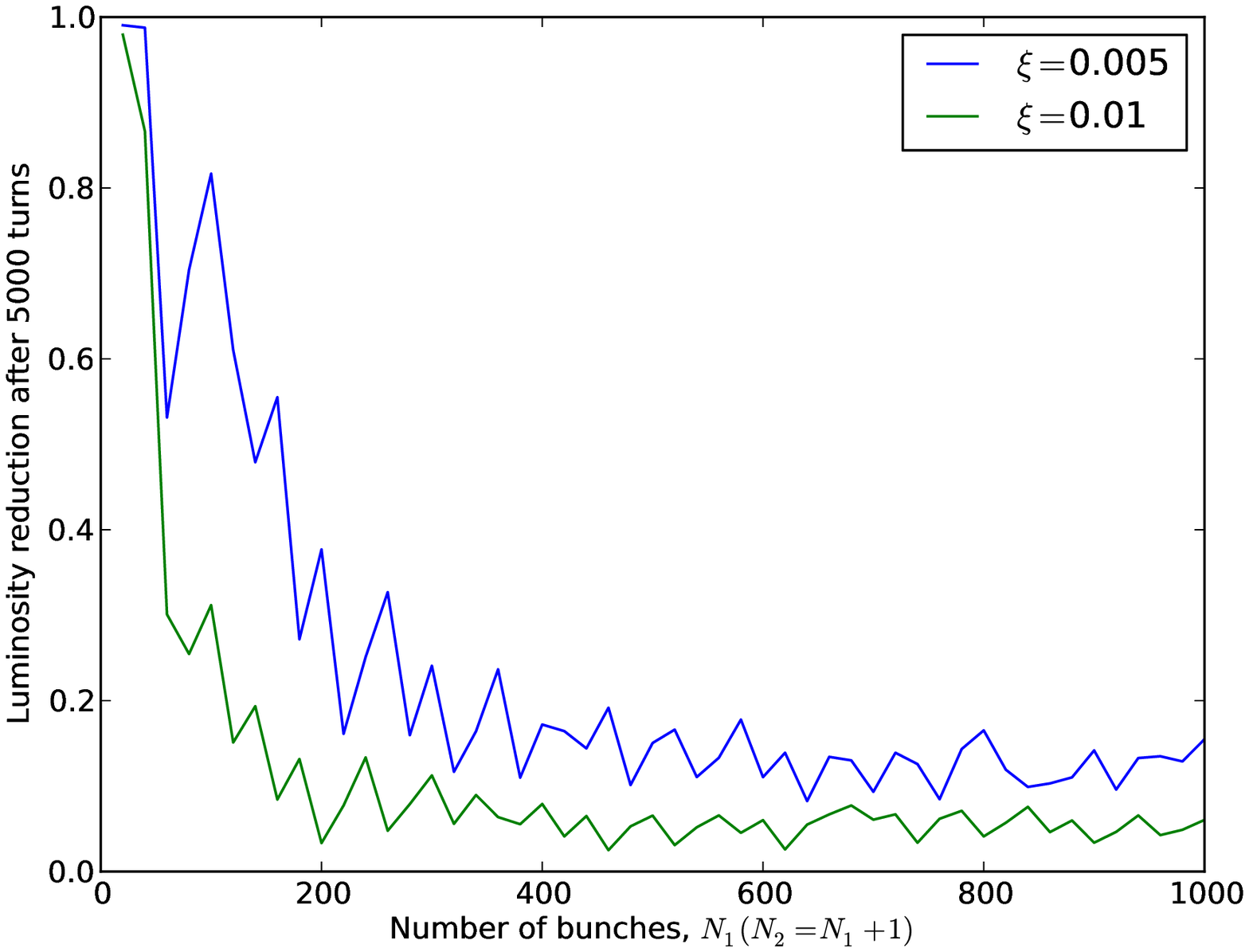}

\caption{Luminosity reduction as function of the bunch number $N_{1}$. \label{fig:lumi-N}}
\end{figure}

Figure \ref{fig:threshold_color_map-quad} illustrates the color map
of the optimized largest modulus of the one-turn matrix $M_{T}$.
Both tunes are optimized in the vicinity of $\nu_{1,2}\sim0.2$. The
stable boundary also is a straight line in the logarithmic scale.
We use the linear regression to obtain the stable boundary condition
as $\xi_{max}\approx0.2N_{1}$, which is, as expected, a tighter condition
than the stable condition for the dipole offset case.

On this linearized model, we imposed two strong assumptions. First
is that the model uses the perturbed vector of the second order momentum.
Therefore even if one-turn $M_{T}$ is unstable, we only can conclude
that the system will deviate from the unperturbed second order momentum
of the beam, which, in a nonlinear system, does not necessarily lead
to instability. Second, we assume the emittance of the beam is constant
after the nonlinear beam-beam interaction and the beam will maintain
a Gaussian distribution with different rms value. However, the emittance
and the transverse distribution of the beam will change slowly under
the nonlinear force especially when beam sizes of the two opposing
beams do not match. Also in our linear model, we overestimated focusing/defocusing
effect of the the beam-beam force, since the particles with larger
amplitude (beyond $2\sigma$ in beam size) experience much smaller
focusing/defocusing force than a linearized one used in the model.
Therefore we are expecting the the stabilization condition in a nonlinear
simulation will be relaxed than that predicted by this linear model.
Therefore, we expect, after scaled by a factor($\leq1$), the maximum
stable beam-beam parameter found in a nonlinear simulation can compare
with that of linear model.

We test the understanding of the accuracy of this model by a multi-bunch
1-D simulation including nonlinear beam-beam force. In a 1-D model,
the coordinates are denoted as $\left(r,r^{\prime}\right)$. The beam-beam
interaction is represented by

\[
\Delta r^{\prime}=\mbox{\ensuremath{\frac{8\pi\xi}{\beta^{*}}}\ensuremath{\ensuremath{\frac{\sigma_{r,0}^{2}}{r}\left(1-\exp\left(-\frac{-r^{2}}{2\sigma_{r}^{2}}\right)\right)}}}
\]
where $\xi$ is the designed beam-beam parameter, $\sigma_{r,0}$
is the initial beam size, which is same for all bunches in both beams
and $\sigma_{r}$ is the rms beam size of the opposing beam before
the beam-beam interaction of each turn. In this simulation, the initial
beam size is set to 100 microns and $\beta^{*}$ is 0.5 meters. We
excluded the dipole effect to ensure that the observation is focused
on the changes in beam size of both beams. We set the number of bunches
in two rings as $\left(N_{1},N_{2}\right)=\left(50,51\right)$. Figure
\ref{fig:quad_simulation} shows the results and compares them with
the linear model. The top left figure shows the instability of multi-bunch
beam size with the beam-beam parameter 0.015. The top right one shows
the luminosity reductions due to this instability at different $\xi$.
We may observe losses as low as $\xi=0.003$, while the system has
returned to a stable condition at $\xi=0.007$ and becomes unstable
again at $\xi=0.01$. We then scanned the beam-beam parameter in the
simulation and compare the finding with the linear matrix model. The
model predicts the result very well except that we had to scale the
beam-beam parameter in the simulation with factor of 0.5 to achieve
a better match. This disparity reflects our over-estimation of the
beam-beam focusing effect in the linear model.

Now, we fix the working point of both rings as 0.68, and vary the
number of bunches in ring 1($N_{1}$) from 20 to 1000, while keeping
$N_{2}=N_{1}+1$. Figure \ref{fig:lumi-N} shows the luminosity reduction
after 5000 turns as function of $N_{1}$. The luminosity loss tends
to saturate at very low level, \textasciitilde{}20 times reduction
at $\xi=0.01$ and \textasciitilde{}8 times reduction at $\xi=0.005$,
when the number of bunches is beyond 400. The zig-zag scattered points
in this figure reflects that the fixed tune (0.68) is not optimized
for different $N_{1}$. In this tune, the resonance strength is larger
for some bunch number, while smaller for others $N_{1}$. The tune
can be optimized by range of $1/N_{1}$ to minimized the unwanted
resonances.

When the system becomes unstable, the luminosity drops quickly in
the first few thousand turns, then stabilize itself after that. The
stabilization results from both the inverse quadratic relation of
the beam-beam parameter as function of the beam size and the nonlinearity
of the beam-beam field. In addition, we observe a slow luminosity
loss (as shown in the bottom right of figure \ref{fig:quad_simulation})
after the self-stabilization. The loss is not observed in a symmetric
scheme, $N_{1}:N_{2}=1:1$, with same beam-beam parameter. We use
linear regression to obtain the speed of the luminosity loss as 2\%
every 100K turns for this beam-beam parameter. For a 3000 meters circumference
ring, the decline in luminosity will be 2\% per second. For an electron-ion
collider, the synchrotron radiation damping of the electron beam may
suppress this slow loss. However, the hadron collider requires aggressive
cooling technology to prevent this slow loss.

In summary, we used similar matrix method to predict an instability
of the second order momentum of the colliding bunches due to the asymmetric
collision pattern. The 1-D simulation confirms the prediction of the
linear model that this instability prevents the choice of reasonable
beam-beam parameter. This instability will entail a very rapid luminosity
loss in milliseconds scale and cannot be overcome by a transverse
damper as the dipole offset case in the previous section. Then the
system is self-stabilized at a much lower luminosity than the design
value and continues suffer from a slow luminosity loss. It is worthwhile
to note that for same tune of ring 2, the effective tunes from $M_{T}$
are generally different in the dipole and quadrupole cases, therefore,
it is impossible to find an optimized tune for both effects.

\[
\]

\section{Single bunch effect with bunch gap\label{sec:bunch_gap}}

Finally, we studied the beam-beam effects in the presence of the bunch
gap. In either ion or electron rings, some buckets have to be empty
for various reasons such as the injection gap, eliminating ion trapping
or electron cloud effects. We assume that the gap is in ring 2, which
can hold $N_{2}$ bunches and that all gaps are positioned together,
i.e. in the $N_{2}$ collisions, the bunch in ring 1 will meet $m$
opposing bunches and $n=N_{2}-m$ empty buckets. The linear matrix
which includes the coherent beam-beam effect of $N_{2}$ turns becomes:

\[
M_{t}=M_{\beta}\left(n\phi\right)\left(M_{k}\left(\xi\right)M_{\beta}\left(\phi\right)\right)^{m}
\]
 where $M_{\beta}$ is the transverse map with one turn phase $\phi$
and $M_{k}$ is the matrix that represent the linearized beam-beam
kick.

\begin{figure}
\includegraphics[width=0.5\columnwidth]{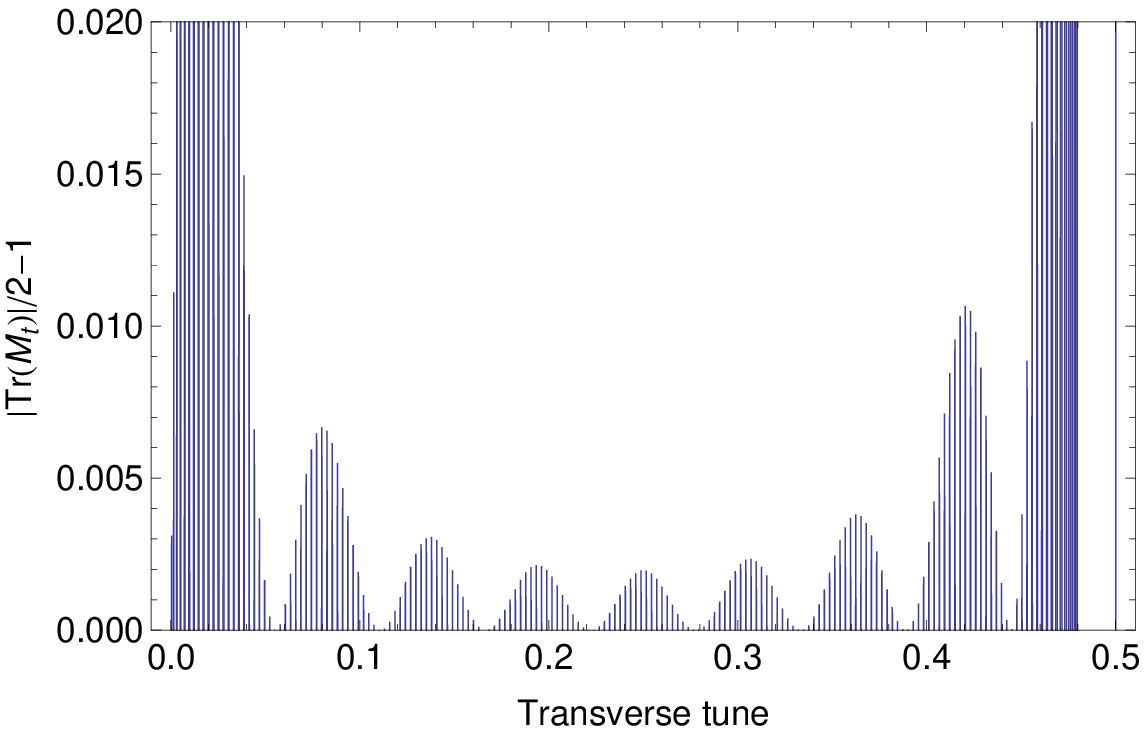}\includegraphics[width=0.5\columnwidth]{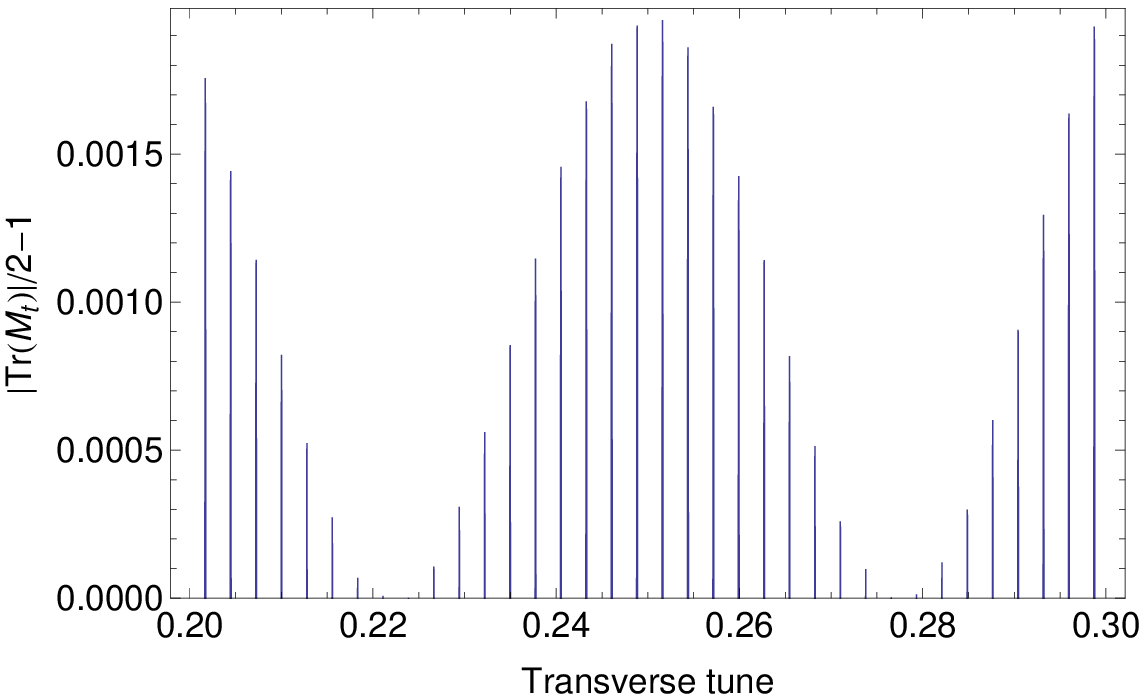}

\includegraphics[width=0.5\columnwidth]{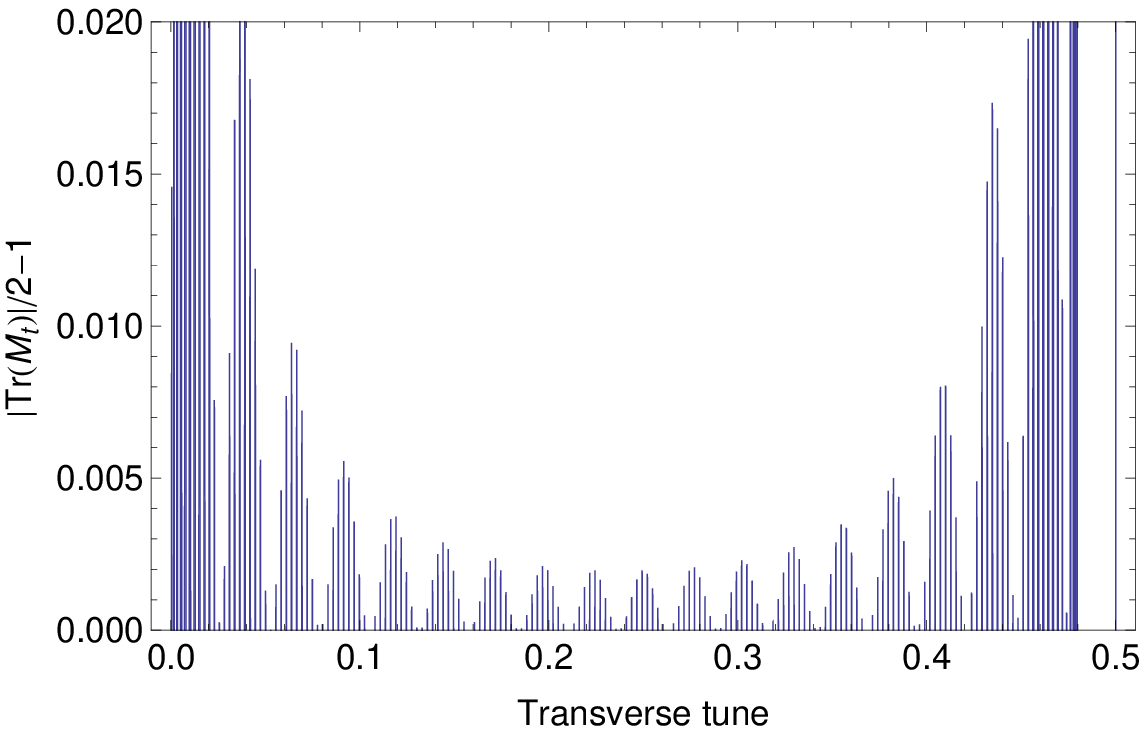}\includegraphics[width=0.5\columnwidth]{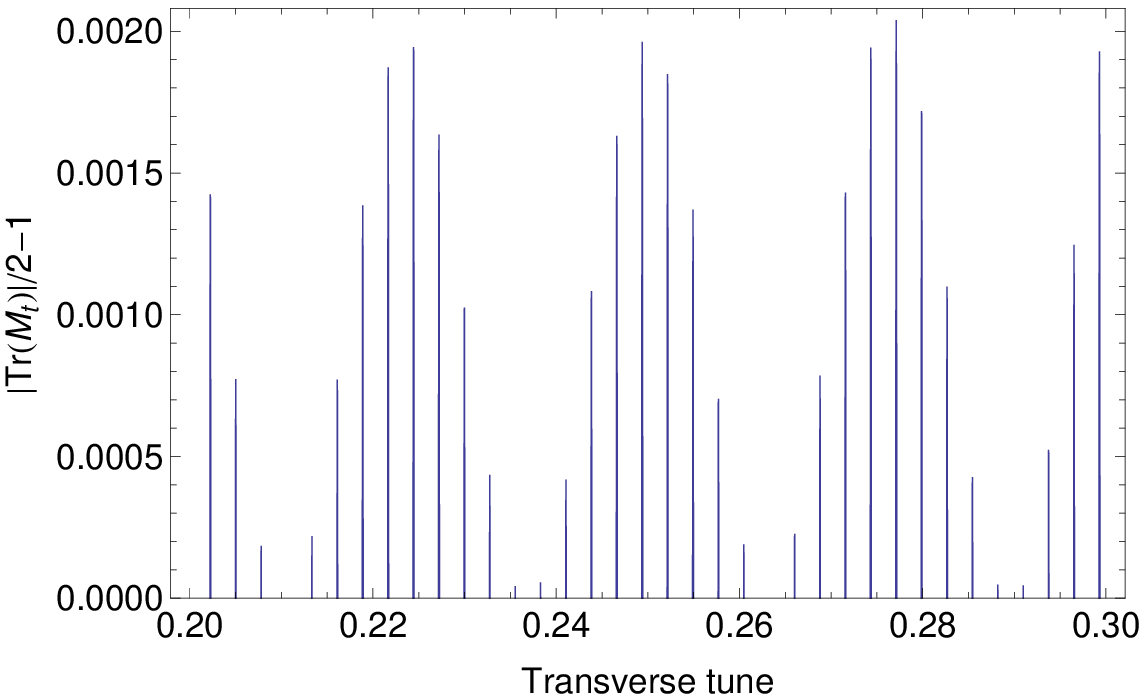}

\caption{$\left|Tr\left(M_{t}\right)\right|/2-1$ as function of betatron tune.
The parameters used in calculation: The beam-beam parameter is 0.01;
$N_{2}=180$. In the top 2 figures, $m=171$ and $n=9$; in the bottom
two figures, $m=161$ and $n=19$. \label{fig: trace of single bunch}
The right figures are the enlarged version of the left-hand ones.}
\end{figure}

The trace of $M_{t}$ determines the beam's stability with a linearized
beam-beam interaction. Figure \ref{fig: trace of single bunch} plots
the $\left|Tr\left(M_{t}\right)\right|/2-1$ , which characterizes
the unstable system via positive values. The result suggest that the
matrix is a unstable one at many specific value of tunes with an approximate
separation of $1/N_{2}$. Those unstable tunes and their amplitude
forms a oscillating envelopes. The valleys between peaks are the largest
stable regions. The number of the envelope peaks equals the number
of missing bunches, $n$.

\begin{figure}
\includegraphics[width=1\columnwidth]{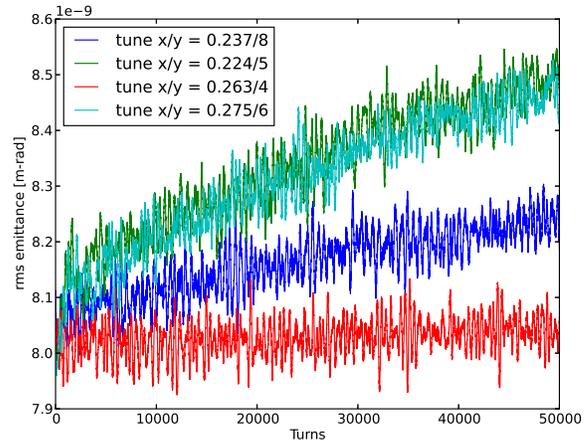}

\caption{Nonlinear 4D beam-beam simulation for different working points. The
following parameters were used in calculation: the beam-beam parameter
is 0.01; $N_{2}=180$; $m=161$ and $n=19$. \label{fig:Nonlinear-4D-beam-beam}}
\end{figure}

In detailing the linear beam dynamics, we always can find a proper
working point that gives stable $N_{2}$ turn matrix $M_{t}$. However,
the nonlinearity of the beam-beam force complicates our working point
optimization, because the tune spread induced by the nonlinear force
may cause the partial particles fall in the unstable region. We used
a 4D nonlinear beam-beam map to simulate the example depicted in figure
\ref{fig: trace of single bunch}. We also included chromaticity in
both ring of unit 1, assuming the rms energy spread was $3\times10^{-4}$.
Figure \ref{fig:Nonlinear-4D-beam-beam} shows the emittance of such
bunch as function of turns. The working points were chosen near the
peaks or valleys as in our study of linear model (bottom right of
figure \ref{fig: trace of single bunch}). With the on-peak working
points (the cyan and the green curve), the growth rate of the beam
emittance is the highest, approximately 5\% every 500 thousand turns.
When the working points are chosen to be in the valleys (the stable
region in linear study), the emittance growth is reduced dramatically.
In this example, the emittance growth is not observed in the first
500 thousand turns when the working points are close to 0.263.

We conclude that in the asymmetric collision scheme with the presence
of gaps in the bunch train, the single bunch has very different stable
condition from that in symmetric case, where the bunch only collide
with one opposing bunch. The choice of working points is limited to
several small regions that defines by the number of gap rf buckets
($n$ in the above study). There will be only $n-1$ stable region
that can be considered. The effect implies another import limitation
on the tune selection of both rings of the collider.

\section{Summary}

We demonstrated the effective 'one-turn' matrix with a permutation
matrix and illustrated the formation of the multi-bunch resonance
and the instability of both beam offsets and beam sizes in a asymmetric
collision pattern when the number of bunches in two beam are different.
We also evaluated the effect of presence of the bunch gap. 

This collision pattern induces new resonances that does not occur
in symmetric case. The separation of the resonance lines are approximately
the reciprocal of the number of bunches in the ring, a feature that
imposes a strong limitation on the stable beam-beam parameter even
after proper working point selection, and therefore restricts the
achievable luminosity. The countermeasure to the dipole offset resonance
is to incorporate an ultra-high gain bunch-by-bunch transverse damper,
which is a very challenging issue, even with state-of-art technology.
The resonance and instability of the beam size of the colliding bunches
will induce fast luminosity loss, and there are have no obvious remedies.
If there is gap in the bunch train, the linear map for a single particle
may become unstable, since the presence of $n$ continuous bunch gaps
will create $n$ unstable regions of the tune space of the ring. The
optimization of tune has different requirements for all 3 effects
in the above discussion, which make it almost impossible to find the
proper value.

We conclude from our studies that this asymmetric collision scheme
('gear-changing' scheme) would introduce a major loss of the luminosity
in colliders and hence should be avoided. We did not identify any
known remedies for this decremental effects.

\begin{multline*}
\\
\end{multline*}

\[
\]


\begin{thebibliography}{9}

\bibitem{eg_rhic}

For example, the Relativistic Heavy Ion Collider (RHIC) at Brookhaven
National Laboratory

\bibitem{eg_elic}

For example, the Electron Light Ion Collider (ELIC) proposed by Jefferson
Laboratory

\bibitem{barycentre_motion}

Kohji Hirata and Eberhard Keil, NIM A, 292 (1990) no.1, 156-168

\end{thebibliography}
\end{document}